\documentclass[aps,twoside]{revtex4}

\begin{document}
\thispagestyle{empty}
\parbox{\textwidth}{\hfill to appear in \emph{Physica A}}
\\
\begin{center}
{\large
Reply to the paper \\[2mm]
E.G.D. Cohen, L. Rondoni: \emph{Particles, Maps and Irreversible Thermodynamics I.}}\\[2mm]
\emph{Physica} \textbf{A 306} (2002) 117
%
\end{center}

In \cite{CohRon02} Cohen and Rondoni severely question the physical
relevance of studies of transport by means of multibaker (MB) maps.
The aim of MB studies (as well as of other work on low-dimensional
chaotic models for transport) is to improve the understanding of
irreversible macroscopic behavior in transport processes along one
spatial dimension. These studies provide an attempt to establish a
macroscopic description consistent with irreversible thermodynamics
(IT) in systems with an \emph{explicit} microscopic dynamics. In this
class of analytically accessible model systems the properties of the
microscopic and macroscopic motion are fully understood such that
their mutual relation can be studied in detail. We believe that it is
admissible to abandon the use of certain concepts of traditional
approaches in order to highlight the general features of irreversible
processes.

The fundamental interest in MB studies lies in identifying minimal
assumptions on the dynamics and on the form of entropy functionals
that lead to a description of the evolution of macroscopic fields
according to the laws of IT --- in particular, to the transport
equations and to the local entropy balance. From this perspective the
arguments of Cohen and Rondoni concerning the assumptions adopted in
the setup of multibaker models are contradictory and part of them do
not hold:

\begin{itemize}
\item %
  Rondoni and Cohen \cite[Sect IV]{CohRon02} proclaim that %
  '{\sffamily it is the lack of real particles that makes local
    thermodynamic equilibrium (LTE) impossible}'.
  We disagree with this statement since LTE only ascertains that the
  thermodynamic description of an equilibrium system applies
  \emph{locally} in sufficiently small volumes of a non-equilibrium
  system. Since thermodynamics is a phenomenological theory, however,
  it does not involve the concept of particles. By definition of its
  basic objects the notion of particles can not be needed for its
  foundation.  Hamiltonian equations of motion or kinetic theory can
  be used to underpin it in certain (dilute gas) limits, but IT is
  more general and can even describe the evolution of systems with a
  different underlying dynamics.
  
  Similar to the Lorentz gas the time evolution of an ensemble of
  trajectories in a multibaker map can be interpreted as an assembly
  of non-interacting particles. By adopting a Boltzmann-Grad type
  limit one can then concentrate on the evolution of fields
  characterizing the state of the system. The characterization of
  these fields forms the prime objective of studies on multibaker
  maps. 
  In the absence of coarse-graining the fields typically develop
  structures on ever refining scales as times evolves. Since their
  evolution is reversible and they never stop evolving, the
  fields are viewed as characteristics of a microscopic motion.
  On the other hand, the averages of the fields over MB cells of
  linear extension $a$ in the transport direction (called
  coarse-grained quantities in what follows) are thermodynamically
  well behaved.  Appropriate averages can be interpreted as density
  $\varrho$, temperature $T$, and hydrodynamic velocity $v$. They are
  considered as local thermodynamic variables, since in a well-defined
  continuum limit their time evolution \emph{exactly} follows the
  transport equations (the advection-diffusion equation, the heat
  conduction equation, and the Navier-Stokes equation of strongly
  viscous fluids) of IT.  Consequently, the cell size $a$ is
  considered as the mesoscopic scale addressed by LTE, below which
  macroscopic concepts do not apply.

  Information-theoretic arguments are evoked to work out the entropy
  balance for a coarse-grained entropy $S$ taking the form
  \begin{equation}
     S = - k_B \varrho \ln\left( \varrho\; T^{-\gamma} \right),
     \label{eq:S}
  \end{equation}
  where $k_B$ is the Boltzmann constant, and $\gamma$ is a positive
  constant playing the role of a dimensionless specific heat.  We have
  shown for a series of transport problems (electric \cite{EarlyWork}
  and heat conduction, cross effects \cite{Matyas}, and viscous flows
  \cite{MTV:PRE01}) that the local entropy balance of this $S$ is
  exactly of the form known from IT. In view of this, (\ref{eq:S}) can
  be interpreted as an equation of state, which holds in \emph{every
    cell} in spite of the spatial and temporal evolution of $\varrho$
  and $T$. This is in our interpretation the analog of LTE. It has
  nothing to do with the concept of particles and their collisions.

\item %
  We agree with Cohen and Rondoni that the expression (\ref{eq:S})
  adopted for the coarse-grained entropy is not obvious, and that it
  is also not clear a priori how one can identify \emph{heat} for a
  model lacking the concept of work and particles.  In our early
  papers \cite{EarlyWork,Matyas} we therefore avoided the use of the term
  heat, and spoke mainly of entropy and entropy production. 
%
%
  In the visco-thermal MB, on the other hand, a clear relation emerged
  \cite{MTV:PRE01}. After coarse graining, the energy balance
  condition leads to the temperature equation
  \[
  \nu \varrho \partial_x T = \lambda \partial^2_x T + Q
  \]
  with
  \[
  Q= \nu \varrho (\partial_x v)^2,
  \]
  where $\nu$ is the kinematic viscosity (a unique expression of the
  MB parameters) which coincides in the model with the heat diffusion
  coefficient, and $\lambda = \nu \varrho \gamma$ represents the heat
  conductivity.  The source term $Q=\nu \varrho (\partial_x v)^2$
  turns out to be the same as the contribution to the entropy
  production due to viscosity multiplied by the temperature. The full
  entropy production takes the form
  \[
  \sigma^{(irr)} = \lambda \left( \frac{\partial_x T}{T} \right)^2
  + \frac{\nu \varrho}{T} (\partial_x v)^2.
  \]
  This strong analogy with IT encouraged us to call $Q$ the viscous
  heating term, since it indeed raises the temperature due to the
  local hydrodynamic momentum exchange of neighboring fluid layers,
  modelled by neighboring MB cells.

\item %
  Concerning the purely diffusive dynamics of the Lorentz gas or an
  isothermal MB,
  which can both be considered as particle models of transport,
  Rondoni and Cohen state that their entropy production only
  \emph{formally} agrees (in an appropriate limit) with the
  thermodynamic expression
  \[
  \sigma^{(irr)} = D \frac{(\nabla \varrho)^2}{\varrho}
  \]
  since the entropy production cannot be related to any heating
  process in the absence of interacting particles 
  (in their words '{\sffamily ...\ in IT the entropy production
    $\sigma_s^{IT}>0$ is due to heat, which involves randomization,
    ...}').
  We mention that IT does not predict $\sigma^{(irr)}$ to be a source
  term of heat in the temperature equation.  Rather [see \cite{dGM}
  chapter XI, eq. (235)], there is no heat source at all in this
  equation, and local changes of the temperature are due to the Dufour
  effect.  The assumtion of constant temperature in the diffusive
  multibaker only implies a vanishing Dufour coefficient.  The entropy
  production $\sigma^{(irr)}$ is in this case fully due to mixing
  entropy induced by the gradient of the macroscopic density
  (phase-space density in the particle picture), which is also present
  in microscopic models of non-interacting particles.

\item %
  The studies of MB models provide a self-consistent framework for
  the description of transport based on a reversible dynamics. In our
  eyes they clearly show that the coarse graining
  carried out over the underlying chaotic and mixing MB dynamics
  unavoidably leads to information loss, which appears to be the basic
  ingredient of irreversible entropy production \emph{and} viscous
  heating. In contrast to the main assertion of \cite{CohRon02} they
  therefore provide a valuable alternative approach to study the
  foundations of IT.
\end{itemize}

Finally, we would like to point out that our interpretation of LTE is
fully consistent with the definition given by the authors of
\cite{CohRon02}, who say:
\\[2mm]
\hspace*{0.05\textwidth} \parbox{0.9\textwidth}{{\sffamily The
    validity of LTE can only be justified \emph{a posteriori}, on the
    grounds of the conclusions derived from it.}}
\\[2mm]
We consider the recovery of the relations of IT from such a simple
model as the MB map as a clear justification for considering 
the cell-averaged, coarse-grained  fields as hallmarks of LTE,
and therefore reject the criticism by Cohen and Rondoni.


\ \\[10mm]
\begin{tabular}{lll}
\parbox{0.36\textwidth}{
L\'aszl\'{o} M\'aty\'as\\
Max-Planck-Institute for the\\ Physics of Complex Systems\\
N\"{o}thnitzer Str.\ 38\\
01187 Dresden\\
Germany}
&
\parbox{0.36\textwidth}{
Tam\'as T\'{e}l\\
Institute for Theoretical Physics\\
E\"{o}tv\"{o}s University\\
P. O. Box 32\\
H-1518 Budapest\\ 
Hungary}
&
\parbox{0.3\textwidth}{
J\"{u}rgen Vollmer\\
Max-Planck-Institute for\\ Polymer Research\\
Ackermannweg 10\\
55128 Mainz\\
Germany}
\end{tabular}
\end{document}